 \documentstyle[amssymb]{article}

\font\mybb=msbm10 at 12pt
\def\bb#1{\hbox{\mybb#1}}
\def\Z {\bb{Z}}
\def\R {\bb{R}}
\def\II {\bb{I}}

\begin{document}
\newcommand{\n}{\nu}
\newcommand{\m}{\mu}
%
%
\renewcommand{\d}{\mbox{$\partial$}}
\newcommand{\bd}{\mbox{$\bar{\partial}$}}
\newcommand{\M}{\mbox{${\cal M}$}}
\newcommand{\F}{\mbox{${\cal F}$}}
\renewcommand{\H}{\mbox{${\cal H}$}}
\renewcommand{\L}{\mbox{${\cal L}$}}
\newcommand{\G}{\mbox{$\hat{G}$}}
\newcommand{\q}{\mbox{${\cal Q}$}}
\newcommand{\tq}{\mbox{$\tilde{\q}$}}
%
%
\newcommand{\s}{\mbox{$\sigma$}}
\newcommand{\f}{\mbox{$\phi$}}
\renewcommand{\O}{\mbox{$\Omega$}}
\newcommand{\x}{\mbox{$\chi$}}
\newcommand{\be}{\begin{equation}}
\newcommand{\ee}{\end{equation}}
\renewcommand{\l}{\mbox{$\lambda$}}
\begin{flushright}
IFT-UAM-97-6\\[.2cm]
hep-th/9712196
\end{flushright}
\vskip .4cm 
\begin{center}
{\Large\bf Black-Holes, Duality and Supersymmetry}\footnotemark{}\\[.5cm]
{\large E. Alvarez$^{\dagger \ddagger}$,
P. Meessen$^{\ddagger}$, 
T. Ort\'{\i}n$^{\dagger\natural}$}\\[.5cm]
{\it $^{\dagger}$ Instituto de F\'{\i}sica Te\'orica, C-XVI, Universidad
Aut\'onoma de Madrid, 28049 Madrid, Spain\\[.2cm]
$^{\ddagger}$ Departamento de F\'{\i}sica Te\'orica, C-XI,
Universidad Aut\'onoma de Madrid, 28049 Madrid, Spain\\[.2cm]
$^{\natural}$ IMAFF, CSIC, Calle de Serrano 121, 28006 Madrid, Spain}\\[.3cm]
{\begin{flushleft}\small
In order to study the discrepancy between the supersymmetry
bound and the extremality bound for rotating black holes,
the effect of duality transformation on the class of
stationary, axially symmetric string backgrounds, called the
TNbh, is considerd.
It is shown that the Bogomol'nyi bound is invariant under those
duality transformations that transform the TNbh into itself, meaning
that duality does not constrain the angular momentum in such a way as
to reconcile the aforementioned bounds. A physical reason for the
existence of the discrepancy is given in terms of superradiance.
\end{flushleft}}
\end{center}
\section{Introduction}
As\footnotetext{Talk given by E. Alvarez at the International
Workshop {\it Beyond the Standard Model: From Theory to Experiment}, 
Val\`encia, October 13-17, 1997.}
is well known, rotating black holes (bh) become extremal
before they become supersymmetric, which is in harrowing
contrast to non-rotating bh's which become supersymmetric
at the same moment they reach extremality, thus giving
support to the Cosmic Censorship Hypothesis \cite{klopp}.
This discrepancy can be summarized by stating that the
angular momentum does enter the extremality bound, typically
in the combination $M^2 -J^2$, $M$ being the mass and $J$ the
angular momentum of the bh, whereas it does not enter the
Bogomol'nyi (supersymmetry) bound.
This discrepancy is even more surprising in
view of the fact that in the presence of only NUT charge (that is, for some
stationary, non-static, cases) both bounds still coincide; the NUT
charge squared must simply be added to the first member in the two
bounds \cite{kall}. On the other hand, it is also known that some
T-duality transformations seem to break spacetime supersymmetry making
it non-manifest \cite{susyt}. These two facts could perhaps give rise
to a scenario in which extremal Kerr-Newman black holes (which are
not supersymmetric) could be dual to some supersymmetric
configuration. At the level of the supersymmetry bounds one would see
the angular momentum transforming under a non-supersymmetry-preserving
duality transformation into a charge that does appear in the
supersymmetry bound (like the NUT charge). In this way, the
constraints imposed by supersymmetry on the charges would
equally constrain the angular momentum.
\par
Although this scenario has been disproven by the
calculations \cite{AMO}\footnote{In fact, the angular momentum is part of
a set of charges that transform amongst themselves under duality
and never appear in the Bogomol'nyi bound.} the transformation of
black-hole charges and the corresponding Bogomol'nyi bounds under
general string duality transformations remains an interesting subject
on its own right and its study should help us gain more insight into
the physical space-time meaning of duality.
\par
In a previous work \cite{quev} a systematic analysis was made of the
behavior of asymptotic charges under T-duality \cite{gipora,aao}
for four-dimensional non-rotating black
holes\footnote{Since some of the objects studied are singular, as
opposed to black holes, the name black hole
will be used in a generalized sense for (usually point-like) objects
described by asymptotics such that a mass, angular momentum etc.~can
be assigned to them.}.
In order to clarify the questions posed in the foregoing paragraphs,
their results are extended, by
essentially widening the class of metrics considered before to
stationary and axially symmetric metrics
and by including more non-trivial
fields, thus enlarging the group of asymptotic-behavior-preserving
transformations.
Therefore, we will define the asymptotic behavior considered
(``TNbh'') and will discuss the subgroup that preserves it (the
``ADS''). One finds that the charges fit into natural multiplets
under the ADS and that the Bogomol'nyi bound can
be written as an invariant of this subgroup. This was to be
expected since duality transformations in general respect unbroken
supersymmetries \cite{susyt}, but since duality in general
transforms conserved charges that appear in the Bogomol'nyi bound into
non-conserved charges (associated to primary scalar hair) that in
principle do not enter, the consistency of the picture will require 
the inclusion of those non-conserved charges into the generalized 
Bogomol'nyi bound.
\section{The Derivation of the Duality Transformations}
\label{sec-derivation}
For simplicity we are going to consider a consistent
truncation
of the four-dimensional heterotic string effective
action, to the lowest order in $\alpha^{\prime}$,
including the metric, dilaton and two-form field plus two
Abelian vector fields. This truncation is, however, rich enough to
contain solutions with $1/4$ of the supersymmetries
unbroken \cite{klopp,ort}.
An extra reason for considering two abelian vector fields, is that
it is the smallest theories, whose solution allows for the
construction, through duality, of the most general bh solution in
$N=4$, $D=4$ Sugra.
The action of interest, in the string frame\footnote{Our
signature is $(-,+,+,+)$. All hatted symbols are
four-dimensional and so $\hat{\mu},\hat{\nu}=0,1,2,3$. The relation
between the four-dimensional Einstein metric
$\hat{g}_{E\hat{\mu}\hat{\nu}}$ and the string-frame metric
$\hat{g}_{\hat{\mu}\hat{\nu}}$ is $\hat{g}_{E\hat{\mu}\hat{\nu}}
=e^{-\hat{\phi}}\hat{g}_{\hat{\mu}\hat{\nu}}$.}, reads 
\begin{equation}
S = \int d^{4}x \sqrt{|\hat{g}|}\ e^{-\hat{\f}} \left[ R(\hat{g})
 +\hat{g}^{\hat{\mu}\hat{\nu}}\d_{\hat{\mu}}\hat{\f}\d_{\hat{\nu}}\hat{\f}
 -{\textstyle\frac{1}{12}} \hat{H}_{\hat{\mu}\hat{\nu}\hat{\rho}}
\hat{H}^{\hat{\mu}\hat{\nu}\hat{\rho}}
-{\textstyle\frac{1}{4}}\hat{F}^{I}{}_{\hat{\mu}\hat{\nu}}
\hat{F}^{I\hat{\mu}\hat{\nu}}
\right]\; ,
\label{s1.1}
\end{equation}
where  $I=1,2 $ sums over the Abelian gauge fields
$\hat{A}^{I}{}_{\hat{\mu}}$ with standard field strengths,
and the two-form field strength is
\begin{equation}
\hat{H}_{\hat{\mu}\hat{\nu}\hat{\rho}}
= 3 \partial_{[\hat{\mu}}\hat{B}_{\hat{\nu}\hat{\rho}]}
-{\textstyle\frac{3}{2}} \hat{F}^{I}{}_{[\hat{\mu}\hat{\nu}}
\hat{A}^{I}{}_{\hat{\rho}]}\, .
\end{equation}
As announced in the Introduction, it will be assumed that the metric
has a timelike and a rotational spacelike isometry.
The former is
physically associated to the stationary
character of the spacetime and the latter to the axial
symmetry.
Since they commute one
can find two adapted coordinates, in this case the time $t$ and the angular
variable $\varphi$, such that the background does not
depend on them.  This then implies that the theory can be
reduced dimensionally.  Using the standard technique \cite{ss,ms} the
resulting dimensionally reduced, Euclidean, action turns out to
be\footnote{Following \cite{sen3} we have set the resulting 2-dimensional
vector fields to zero. Other choices could lead to a 2 dimensional
cosmological constant.}
\begin{eqnarray}
S & = & \int d^{2}x \sqrt{|g|}\ e^{-\f}
\left[  R(g)+g^{\mu\nu}\d_{\mu}\f\d_{\nu}\f
  +{\textstyle\frac{1}{8}}{\rm Tr}\d_{\mu}\M\d^{\mu}\M^{-1}
\right. \nonumber \label{s1.2} \\
& &
\left.
-{\textstyle\frac{1}{4}}W^{i}_{\mu\nu}(\M^{-1})_{ij}W^{j\mu\nu}\right]\; ,
\nonumber
\end{eqnarray}
where $\mu\nu=2,3$ and
$\alpha, \beta=0,1$. The two-dimensional
fields are the metric $g_{\mu\nu}$, six vector fields ${\cal
  K}^{i}{}_{\mu} = (K^{(1)\alpha}{}_{\mu}, K^{(2)}{}_{\alpha\mu},
K^{(3)I}{}_{\mu})$ with the standard Abelian field strengths
$W^{i}{}_{\mu\nu}$ ($i=1,\ldots,6$) and a bunch of scalars
$G_{\alpha\beta},\hat{B}_{\alpha\beta},\hat{A}^{I}{}_{\alpha}$ that
appear combined in the $6\times 6$ matrix ${\cal M}_{ij}$. They are
given by
\begin{equation}
\begin{array}{lcllcl}
G_{\alpha\beta} & = & \hat{g}_{\alpha\beta}\, , &
\phi & = & \hat{\phi}  -{\textstyle\frac{1}{2}}\log \mid\det
G_{\alpha\beta}\mid \, , \\
K^{(1)\alpha}{}_{\mu} & = &
\hat{g}_{\mu\beta} (G^{-1})^{\beta\alpha}\, , &
C_{\alpha\beta} & = & {\textstyle\frac{1}{2}}
\hat{A}^{I}{}_{\alpha}\hat{A}^{I}{}_{\beta} +\hat{B}_{\alpha\beta}\, , \\
g_{\mu\nu} & = & \hat{g}_{\mu\nu}
-K^{(1)\alpha}{}_{\mu}K^{(1)\beta}{}_{\nu}G_{\alpha\beta}\, , &
K^{(3)I}_{\mu} & = &
\hat{A}^{I}{}_{\mu} -\hat{A}^{I}{}_{\alpha} K^{(1)\alpha}{}_{\mu}\, , \\
K^{(2)}{}_{\alpha\ \mu} & = & \hat{B}_{\mu\alpha}
+\hat{B}_{\alpha\beta} K^{(1)\beta}{}_{\mu}
+{\textstyle\frac{1}{2}} \hat{A}^{I}{}_{\alpha}K^{(3)I}{}_{\mu}\, ,
\hspace{-1cm} &
& & \\
\end{array}
\end{equation}
and
\small
\begin{eqnarray}
(\M_{ij}) &=&
\left(
\begin{array}{ccc}
G^{-1}& -G^{-1}C & -G^{-1}A^{T} \\
& & \\
-C^{T}G^{-1} & G+C^{T}G^{-1}C +A^{T}A &
C^{T}{G^{-1}} A^{T} +A^{T} \\
& & \\
-AG^{-1} & AG^{-1}C+A & \II_{2} + AG^{-1} A^{T} \\
\end{array}
\right)\, ,
\label{s1.3}
\end{eqnarray}
\normalsize
$A$ being the $2\times 2$ matrix with entries
$\hat{A}^{I}{}_{\alpha}$. If $B$ stands for the $2\times 2$ scalar
matrix $(\hat{B}_{\alpha\beta})$, then the $2\times 2$ scalar matrix
$C$ is given by
\begin{equation}
C = {\textstyle\frac{1}{2}} A^{T}A + B\, .
\end{equation}
%
The matrix $\M$ satisfies $\M\L\M\L = \II_{6}$, with
\begin{equation}
 \L \equiv
\left(
\begin{array}{ccc}
0& \II_{2} & 0 \\
\II_{2}& 0 & 0 \\
0& 0& \II_{2}
\end{array}
\right)\; .
\label{s1.4}
\end{equation}
From Eq.~(\ref{s1.2}) it is obvious that the dimensionally
reduced action, is invariant under the global transformations given by
\begin{equation}
\M \rightarrow \O\ \M\ \O^{T} \; ,
\hspace{1.5cm}
{\cal K}^{i}{}_{\mu} \rightarrow {\O^{i}}_{j}\ {\cal K}^{j}{}_{\mu}\, ,
\label{s1.5}
\end{equation}
iff the transformation matrices $\Omega$ satisfy the identity
$\Omega\ {\cal L}\ \Omega^{T}={\cal L}$,
showing that the action is invariant under $O(2,4)$, the
classical duality group.
%
\par
The $N=4,d=4$ supergravity equations of motion are actually invariant
under S-duality \cite{kn:CSF}, although this isn't obvious from
Eq. (\ref{s1.1}). In order to write Eq. (\ref{s1.1}) in a manifest
S-duality invariant form one goes over to the Einstein frame and
Hodge dualizes the 3-form $\hat{H}$, imposing that it satisfies
the Bianchi identity, so that the the number of
degrees of freedom is not changed. The action then reads
\begin{equation}
\label{sdual3}
S= \int d^{4}x\sqrt{|g_{\scriptstyle E}|} \left[ \hat{R}(\hat{g}_{E})
-{\textstyle\frac{1}{2}}\left(\Im{\rm m}\hat{\lambda}\right)^{-2}
\partial_{\hat{\mu}}\hat{\lambda}
\partial^{\hat{\mu}}\bar{\hat{\lambda}}
+{\textstyle\frac{1}{4}}\hat{F}^{I}\, {}^{\star}
\tilde{\hat{F}^{I}} \right]\, .
\end{equation}
where we have defined
\begin{equation}
\hat{\lambda} \equiv \hat{a} +ie^{-\hat{\phi}}\, ,
\tilde{\hat{F}^{I}}\equiv e^{-\hat{\phi}}\ {}^{\star}\hat{F}^{I}
+\hat{a}\hat{F}^{I} \, ,
\partial_{\hat{\mu}}\hat{a} = \frac{1}{3!\sqrt{|\hat{g}_{E}|}}
e^{-2\hat{\phi}}\hat{\epsilon}_{\hat{\mu}\hat{\nu}\hat{\rho}\hat{\sigma}}
\hat{H}^{\hat{\nu}\hat{\rho}\hat{\sigma}}\, .
\label{axdef.1}\end{equation}
The S-duality transformations then take the form
\begin{equation}
\hat{\lambda}^{\prime} = \frac{a\hat{\lambda}+b}{c\hat{\lambda}+d}\, ,\;
\left(\begin{array}{c} \tilde{\hat{F}}^{I\prime}\\
		      \hat{F}^{I\prime}\end{array}\right) \;=\;
\left(\begin{array}{cc} a & b\\ c& d\end{array}\right)\,
\left(\begin{array}{c} \tilde{\hat{F}}^{I}\\ 
\hat{F}^{I}\end{array}\right) \; ,
\end{equation}
with $ad-bc=1$, meaning that S-duality is $Sl(2,\R )$.
$Sl(2,\R)$ is generated by three types of
transformations:
 rescalings
$\hat{\lambda}^{\prime} = a^{2}\hat{\lambda}$,
continuous shifts of the axion
$\hat{\lambda}^{\prime} = \hat{\lambda} +b$,
and the discrete transformation
$\hat{\lambda}^{\prime} = -1/\hat{\lambda}$.
%
%
\section{TNbh Asymptotics}
\label{sec-asymptotic}
The asymptotic behavior of four-dimensional asymptotically flat metrics is
completely characterized to first order in $1/r$ by only two charges:
the ADM mass $M$ and the angular momentum $J$. Duality, however,
transforms asymptotically flat metrics into non-asymptotically flat
metrics which need different additional charges to be characterised
asymptotically.  One of them \cite{quev} is the NUT charge $N$ and
closure under duality forces us to consider it.
With the aforementioned conditions on the four-dimensional metric
it is always possible to choose coordinates such that the
Einstein metric in the $t-\varphi$ subspace has the following
expansion in powers of $1/r$:\footnote{We will only write those
terms in the asymptotic expansion that will actually be used, although
more general terms are possible.}
\small
\begin{eqnarray}
\label{metricchoice}
\left( \hat{g}_{E\alpha\beta}\right)  \hspace{-.3cm} & = &  \hspace{-.3cm}
 \left(
 \begin{array}{lr}
  -1+2M/r  &
 \hspace{-4cm}2N\cos \theta  +2J\sin^{2}\theta /r \\
& \\
2N\cos \theta +2J\sin^{2} \theta /r &
(r^{2}+2Mr)\sin^{2}\theta
\end{array}
\right)
\nonumber
\end{eqnarray}
\normalsize
We will assume the following behavior for the dilaton
\begin{equation}
e^{-\hat{\phi}} = 1-2{\cal Q}_{d}/r
+2 {\cal W}\cos\theta/r^{2} -2{\cal Z}/r^{2}
+{\cal O}(r^{-3})\, ,
\end{equation}
where ${\cal Q}_{d}$ is the dilaton charge and
${\cal W}$ is a charge related to the angular momentum that will be
forced upon us by S-duality.
Observe that we have fixed the constant asymptotic value equal to zero
using the same reasoning as Burgess {\it et.al.} \cite{quev},
i.e.~rescaling it away any time it arises.  The time coordinate, when
appropriate, will be rescaled as well, in order to bring the
transformed Einstein metric to the above form (i.e.~to preserve our
coordinate (gauge) choice), but in a duality-consistent way.
Sometimes it will also be necessary to rescale the angular coordinate
$\varphi$ in order to get a metric looking like (\ref{metricchoice}).
Conical singularities are then generically induced, and then the
metric is not asymptotically TNbh in spite of looking like
(\ref{metricchoice}).
\par
The objects we will consider will generically carry electric (${\cal
  Q}_{e}^{I}$) and magnetic (${\cal Q}_{m}^{I}$) charges with respect
to the Abelian gauge fields $\hat{A}^{I}{}_{\hat{\mu}}$.  Since we
allow also for angular momentum, they will also have electric (${\cal
  P}_{e}^{I}$) and magnetic (${\cal P}_{m}^{I}$) dipole momenta. This
implies for the two-dimensional scalar matrix $A$ the following
asymptotic behavior\footnote{One could include constant parts in
the gauge fields. Usually they can be absorbed by a gauge transformation,
but in doing dimensional reduction the gauge group gets broken, giving
the constant parts an invariant meaning.
See \cite{AMO,quev} for a more elaborated discussion.}
\begin{equation}
\hat{A}^{I}{}_{t}= -2
{\cal Q}^{I}_{e}/r +2{\cal P}_{e}^{I}\cos\theta /r^{2}
\; ,\;
\hat{A}^{I}{}_{\varphi} = -2
     {\cal Q}^{I}_{m}\cos\theta -2{\cal P}_{m}^{I}\sin^{2}\theta /r \; .
\label{s2.2}
\end{equation}
Electric dipole momenta appear at higher order in $1/r$ and it is not
strictly necessary to consider them from the point of view of
T-duality, since it will not interchange them with any of the other
charges we are considering and that appear at lower orders in $1/r$.
However, S-duality will interchange the electric and magnetic dipole
momenta and we cannot in general ignore them\footnote{The
different behavior of T- and S-duality is due to the fact that
T-duality acts on the potential's components whereas S-duality acts on the
field strengths}.
\par
The two-index form will have the usual charge $\q_{a}$.  Closure under
duality again demands the introduction of a new extra parameter
(``charge'') that we denote by $\F$ and which will interchange
with $J$ under duality.
The asymptotic
expansion is, then
\begin{equation}
\hat{B}_{t\varphi} \; =\;
\q_{a}\cos \theta  +\F\sin^{2}\theta/r \; .
\label{s2.3}
\end{equation}
Using the definition in Eq. (\ref{axdef.1}) for the pseudoscalar
$\hat{a}$ we can find its, allowing for a
constant value at infinity $\hat{a}_{0}$ which will be set to zero in
the initial configuration, asymptotic expansion to be
\begin{equation}
\hat{a} = \hat{a}_{0} + 2{\cal Q}_{a}/r -2{\cal F}\cos\theta/r^{2} \; ,
\end{equation}
showing that ${\cal Q}_{a}$ is the standard axion charge \cite{ort}.
\par
The class of asymptotic behavior just described, determined by the
twelve charges
$M,J,N,{\cal Q}_{a},{\cal F},{\cal Q}_{d},
{\cal Q}_{e}^{I},{\cal Q}_{m}^{I},{\cal P}_{m}^{I}$
will henceforth be referred to as {\it TNbh
asymptotics}.
\section{Transformation of the Charges under Duality}
\label{sec-transformations}
As is well known \cite{helgas},
the T-duality group can be decomposed locally as
$O(2,4)\sim SO^{\uparrow}(2,4)\otimes\Z_{2}^{(B)}\otimes\Z_{2}^{(S)}$,
where $SO^{\uparrow}(2,4)$ is the connected component containing the
identity of $O(2,4)$. One can show that $\Z_{2}^{(S)}$ can be taken
to be generated by $\II_{6}$ and $-\II_{6}$, thus acting trivially,
and that $\Z_{2}^{(B)}$ generates \cite{BJO}
the Buscher transformtions \cite{buscher}.
\par
The action of the T-duality group was discussed in \cite{AMO},
where it was shown that only a seven dimensional
subgroup of $SO^{\uparrow}(2,4)$
transforms TNbh into TNbh, the ADS, and that the charges transform
linearly under the ADS.
Actually, the ADS closes on sets of four charges and its effect on
the charges can be described by a four dimensional matrix representation
acting on three charge-vectors:
\begin{equation}
\vec{M} \equiv
(
M ,{\cal Q}_{d} , {\cal Q}_{e}^{1} , {\cal Q}_{e}^{2}
)\, ,
\hspace{.3cm}
\vec{N} \equiv
(
N , {\cal Q}_{a} , {\cal Q}_{m}^{1} , {\cal Q}_{m}^{2}
)\, ,
\hspace{.3cm}
\vec{J} \equiv
(
J , {\cal F} , {\cal P}_{m}^{1} , {\cal P}_{m}^{2}
)\, ,
\label{eq:chargevectors}
\end{equation}
which will be referred to, respectively, as
{\it electric, magnetic} and {\it dipole} charge vectors.
There is a fourth charge vector that contains the electric dipole
momenta ${\cal P}_{e}^{I}$, the dilaton dipole-type charge ${\cal W}$
and an unidentified geometrical charge which we denote by $K$
\begin{equation}
\vec{K}\equiv
(
K , {\cal W} , {\cal P}_{e}^{1} , {\cal P}_{e}^{2}
)\; .
\label{eq:multipletK}
\end{equation}
The presence of this fourth charge vector is required
by S-duality, as will be explained later.
\par
%
%
\par
The generator of $\Z_{2}^{(B)}$ is not unique (it is an element of a
coset group).
Two obvious choices correspond to the Buscher transformations in the
directions $t$ and $\varphi$. The Buscher transformation in the
direction $\varphi$ does not preserve TNbh asymptotics and so we will
take as generator of $\Z_{2}^{(B)}$ the Buscher transformation in the
direction $t$,\footnote{Note that the transformation taking Buscher´s
transformation in the $\varphi$ direction into $\tau$ is
not part of the ADS, which is as it ought to be.}
that we denote by $\tau$, with matrix
\begin{equation}
\Omega_{\eta}(\tau) =
\left(
\begin{array}{cc}
+1 &           \\
   & - \II_{5} \\
\end{array}
\right)\, .
\end{equation}
The effect of $\tau$ on all the charges can be expressed in terms of
the same symmetric $4\times 4$ matrix $\Omega_{\tau}^{(4)}$
\begin{equation}
\Omega_{\tau}^{(4)}=
\left(
\begin{array}{ccc}
0 & 1 &         \\
1 & 0 &         \\
  &   & \II_{2} \\
\end{array}
\right)\, ,
\label{rep.tau}
\end{equation}
acting on the charge vectors $\vec{M},\vec{N},\vec{J}$.
The involutive property, that on the charges $\tau^{2}=id$ is
immediately apparent.
\par
Looking at the action of the TNbh preserving transformations \cite{AMO}
one can see that they act as the group $O(1,2)$ on the charges. This
then not only means that the multiplets get broken up into a
singlets and triplets, but also that T-duality leaves invariant
the four dimensional metric $\eta^{(4)}={\rm diag}(+,+,-,-)$, which, 
as we will
see, will be of some importance whilst discussing the transformations
of the Bogomol'nyi bound under duality.
%
\par
The transformation of the electric, magnetic, dilaton and axion
charges under S-duality has been previously studied in
Ref. \cite{ort,kall2}. Here we are considering more charges and we are
choosing initial configurations with vanishing asymptotic values of
the axion and dilaton. In general, S-duality generates non-vanishing
values of these constants and we will remove them by applying further
S-duality transformations.
The net effect \cite{AMO} is that one allows only an $SO(2)$ subgroup
of S-duality whose
result, expressed in terms of the entries of
the original $SL(2,\R)$ matrix is
\begin{equation}
\left(
\begin{array}{c}
{\cal Q}_{e}^{I\ \prime} \\
\\
{\cal Q}_{m}^{I\ \prime} \\
\end{array}
\right)
=
\left(
\begin{array}{cc}
\frac{d}{\sqrt{c^{2}+d^{2}}} & \frac{c}{\sqrt{c^{2}+d^{2}}} \\
& \\
\frac{-c}{\sqrt{c^{2}+d^{2}}} & \frac{d}{\sqrt{c^{2}+d^{2}}} \\
\end{array}
\right)
\left(
\begin{array}{c}
{\cal Q}_{e}^{I} \\
\\
{\cal Q}_{m}^{I} \\
\end{array}
\right)\, ,
\end{equation}
and similarly for the vector of dipole momenta
$\left({\cal P}_{m}^{I}\, ,{\cal P}_{e}^{I}\right)$ and
\begin{equation}
\left(
\begin{array}{c}
{\cal Q}_{d}^{\prime} \\
\\
{\cal Q}_{a}^{\prime} \\
\end{array}
\right)
=
\left(
\begin{array}{cc}
\frac{d^{2}-c^{2}}{\sqrt{c^{2}+d^{2}}} & \frac{2cd}{\sqrt{c^{2}+d^{2}}} \\
& \\
\frac{-2cd}{\sqrt{c^{2}+d^{2}}} & \frac{d^{2}-c^{2}}{\sqrt{c^{2}+d^{2}}} \\
\end{array}
\right)
\left(
\begin{array}{c}
{\cal Q}_{d} \\
\\
{\cal Q}_{a} \\
\end{array}
\right)\, ,
\end{equation}
and, analogously for the charge vector
$\left({\cal W}\, ,{\cal F}\right)$. Observe that the last $SO(2)$
transformation matrix is precisely the square of the former.
\par
It is now clear that the multiplet structure that we built for the
T-duality transformations is not respected by S-duality: the last
three components of the ``electric'' multiplet $\vec{M}$ are rotated
into the last three components of the ``magnetic'' multiplet $\vec{N}$
and vice versa. The same happens with the multiplet $\vec{K}$ defined
in Eq.~(\ref{eq:multipletK}), whose last three components are rotated
into those of the multiplet $\vec{J}$ in exactly the same way, and
vice versa (this is the reason why we  introduced $K$ and $\vec{K}$ in
the first place). To respect the T-duality multiplet structure and, at
the same time incorporate the S-duality multiplet structure it is useful
to introduce the complexified multiplets
\begin{equation}
\label{eq:calM}
\vec{\cal M} \equiv \vec{M} +i\vec{N} \hspace{.6cm} \mbox{and}
\hspace{.6cm} \vec{\cal J} \equiv \vec{K} +i\vec{J} \; .
\end{equation}
These two complex vectors transform under T-duality with exactly the
same matrices as the real vectors and, under the
above S-duality transformations with the complex matrix
$\Sigma^{(4)}=diag(1, e^{2i\theta}, e^{i\theta}, e^{i\theta})$, with
$\theta= {\rm Arg}(d-ic)$,
so
\begin{equation}
\vec{\cal M}^{\prime} =  \Sigma^{(4)} \vec{\cal M}\, ,
\hspace{1cm}
\vec{\cal J}^{\prime} =  \Sigma^{(4)} \vec{\cal J}\, .
\end{equation}
\section{The Bogomol'nyi Bound and its Variation}
In $N=4$ supergravity there are two Bogomol'nyi (B.) bounds, of the form
\begin{equation}
M^{2}-|Z_{i}|^{2}\geq 0\, ,\hspace{1cm} i = 1,2  \, ,
\end{equation}
where the $Z_{i}$'s are the complex skew eigenvalues of the
central charge matrix and are combinations of electric and magnetic
charges of the six graviphotons. These two bounds can be combined into
a single bound by multiplying them and then dividing by $M^{2}$.
One
then gets a {\it generalized} B. bound
\begin{equation}
M^{2} +\frac{|Z_{1}Z_{2}|}{M^{2}} -|Z_{1}|^{2}- |Z_{2}|^{2}\geq 0\, .
\end{equation}
In regular black-hole solutions the second term can be identified with
scalar charges of ``secondary'' type. The identification is, actually
(with zero value for the dilaton at infinity)
\begin{equation}
\frac{|Z_{1}Z_{2}|}{M^{2}} = {\cal Q}^{2}_{d} +{\cal Q}^{2}_{a}\, .
\end{equation}
%
Inserting then the expressions for the central charges one obtains
a generalized B. bound \cite{ort}, which doesn't take account
of the NUT charge, meaning that the B. bound is valid for asymptotically
flat spaces only.
This problem can be overcome by the
reasoning of Ref. \cite{kall} where it was observed that the NUT charge
$N$ does enter in the generalized B. bound. With our definitions the B.
bound for asymptotically TNbh spaces takes the form
\begin{equation}
 M^{2} +N^{2}
+{\cal Q}_{d}^{2} + {\cal Q}_{a}^{2}
- {\cal Q}_{e}^{I}{\cal Q}_{e}^{I}
-{\cal Q}_{m}^{I}{\cal Q}_{m}^{I}
\geq 0 \, .
\label{bogo1}
\end{equation}
To study the transformation properties of the B. bound under the
physical TNbh asymptotics-preserving duality group it is convenient to
use the diagonal metric of $SO(2,2)$ $\eta^{(4)}={\rm diag}
(1,1,-1,-1)$ already introduced in section (\ref{sec-transformations}).
Using this metric and the charge vectors defined in
Eq.~(\ref{eq:calM}) the B. bound can be easily rewritten
to the form
\begin{equation}
\vec{\cal M}^{\dagger} \eta^{(4)}\vec{\cal M}\geq 0\, .
\end{equation}
In this form the B. bound of $N=4,d=4$ supergravity is manifestly
$U(2,2)$-invariant. Observe that $U(2,2)\sim O(2,4)$, although it is not
clear if this fact is a mere coincidence or it has a special significance.
The T-duality piece of the ADS is an $O(1,2)$ subgroup of the $O(2,2)$
canonically embedded in $U(2,2)$ and obviously preserves the B. bound.
The S-duality piece of the ADS is a $U(1)$ subgroup diagonally embedded
in $U(2,2)$ through the matrices $\Sigma^{(4)}$
and obviously preserve the B. bound.
\section{Conclusions}
\label{sec-conclusions}
In the course of our argument we saw that duality acts linearly
on the charges defining stationary, axially symmetric, possibly
non-asymptotically flat string backgrounds, and that duality
collects these charges in tetraplets, see Eq. (\ref{eq:chargevectors},
\ref{eq:multipletK}).
It was further shown that the charges in the vector $\vec{\cal J}$ do 
not appear in the $D=4$, $N=4$ Bogomol'nyi bound
and that neither T- nor S-duality changes this fact. It is not possible to
constrain the values of any of the charges it (in particular $J$)  by
using duality and supersymmetry, as was suggested in the Introduction.
\par
Although our reasoning is completely clear when we look at specific
solutions one should be able to derive B. bounds including primary
scalar charges using a Nester construction based on the supersymmetry
transformation laws of the fermions of the supergravity theory under
consideration. To be able to do this one has to be able to manage more
general boundary conditions including the seemingly unavoidable naked
singularities that primary hair implies.
\par
The results discussed so far
leave unanswered the question hinted at in the
Introduction: why does the angular momentum appear in the definition of
extremality (defining the borderline between a regular horizon and a
naked singularity, with zero Hawking temperature) but not in the
Bogomol'nyi bound (whose saturation guarantees absence of quantum
corrections, as well as a ``zero force condition'', allowing
superposition of static solutions)?
We hold this to be due to the fact that stationary (as opposed
to static) black holes possess a specific decay mode, known in
the bh literature as ``superradiance" \cite{bdw,kn:FHM}, even visible
classically by scattering waves off black holes.
The way this appears is that the amplitude for reflected waves is
greater than the corresponding incident amplitude, for low
frequencies, up to a given frequency cutoff, $m\Omega_H$, depending on
the angular momentum of the hole, and such that $\Omega_H(a=0)=0$.
The angular momentum of the hole decreases by this mechanism until a
static configuration is reached.  The physics underlying this process
is similar to the one supporting Penrose's energy extraction
mechanism, i.e. the fact that energy can be negative in the
ergosphere. This, in turn, is a straightforward consequence of the
fact that the Energy of a test particle is defined as $E
= p.k$, where $p$ is the momentum of the particle, and $k$ is the
Killing vector (which has spacelike character precisely in the
ergosphere); and the product of a spacelike and a timelike vector
does not have a definite sign.
\par
Quantum mechanically this means that there are two competing mechanisms
of decay for a rotating (stationary) black hole:
spontaneous emission (the quantum effect associated to the superradiance),
which is not thermal (and disappears when the angular momentum goes to zero)
and Hawking radiation, which is thermal.
The first is more efficient for massive black holes, but its width is
never zero until the black hole has lost all its
angular momentum.
\par
This then shows that even if the black hole is extremal, it cannot
be stable quantum mechanically as long as its angular momentum is
different from zero. This argument taken literally would suggest that
it is not possible to have BPS states with non zero angular momentum,
unless they are such that no ergosphere exists. This is the case of
the supersymmetric Kerr-Newman solutions which are singular and,
therefore, do not have ergosphere. What is not clear is why
supersymmetry signals the singular case as special and not the usual
extremal Kerr-Newman black hole.\footnote{This argument seems to be
valid only in four dimensions though, since rotating charged black
holes which are BPS states exist in five dimensions \cite{kn:BMPV}.
The existence of two Casmirs for the five-dimensional angular
momentum seems to play an important role.}
%
\section*{Acknowledgments}
The work has been partially suported by the spanish grants AEN/96/1655
(E.A., T.O.), AEN/96/1664 (E.A.) and by the european grants
FMBI-CT-96-0616 (P.M.), FMRX-CT96-0012 (E.A., T.O.).
T.O. would like to thank M.M. Fern\'andez for her support.
E.A. would like to thank the organizers for their invitation and
hotspitality.

%

\section*{References}
\begin{thebibliography}{meessen71}
\bibitem{klopp} R.~Kallosh, A.~Linde, T.~Ort\'{\i}n, A.~Peet,
                A.~Van Proeyen;
                {\it Phys.~Rev.} {\bf D46}(1992), 5278.
\bibitem{kall} R.~Kallosh, D.~Kastor, T.~Ort\'{\i}n, T.~Torma:
               {\it Phys.~Rev.}~{\bf D50}(1994), 6374.
\bibitem{susyt} I.~Bakas:
                {\it Phys.~Lett.}~{\bf 343B}(1995), 103.\\
                E.~Bergshoeff, R.~Kallosh, T.~Ort\'{\i}n:
                {\it Phys.~Rev.}~{\bf D51}(1995), 3009.\\
                I.~Bakas, K.~Sfetsos:
                {\it Phys.~Lett.}~{\bf 349B}(1995), 448.\\
                S.F.~Hassan:
                {\it Nucl.~Phys.}~{\bf B460}(1996), 362.\\
                E.~\'Alvarez, L.~\'Alvarez-Gaum\'e, I.~Bakas:
                {\it Nucl.~Phys.}~{\bf B457}(1995), 3;
                {\it Nucl.~Phys.~Proc.~Suppl.}~{\bf 46}(1996), 16.
\bibitem{AMO} E. Alvarez, P. Meessen, T. Ort\'{\i}n:
               {\sl Transformation of Black-Hole Hair under Duality
                 and Supersymmetry}, preprint FTUAM-97-4, CERN-TH/97-77
               and {\bf hep-th/9705094}.
\bibitem{quev} C.P.~Burgess, R.C.~Myers, F.~Quevedo:
               {\it Nucl.~Phys.}~{\bf B442}(1995), 97.
\bibitem{gipora} A.~Giveon, M.~Porrati, E.~Rabinovici:
                 {\it Phys.~Rep.}~{\bf 244}(1994), 77.
\bibitem{aao} E.~\'Alvarez, L.~\'Alvarez-Gaum\'e, Y.~Lozano:
              Nucl. Phys. (Proc. Suppl.) {\bf 41}(1995), 1.
\bibitem{ort} T.~Ort\'{\i}n:
              {\it Phys.~Rev.}~{\bf D47}(1993), 3136;
              {\it Phys.~Rev.}~{\bf D51}(1995), 790.
\bibitem{ss} J.~Scherk, J.H.~Schwarz:
             {\it Nucl.~Phys.}~{\bf B153}(1979), 61.
\bibitem{ms} J.~Maharana, J.H.~Schwarz:
             {\it Nucl.~Phys.}~{\bf B390}(1993), 3.
 \bibitem{sen3}  A.~Sen:
                 {\it Nucl.~Phys.}~{\bf B447}(1995), 62.
\bibitem{kn:CSF} E.~Cremmer, J.~Scherk, S.~Ferrara:
                 {\it Phys.~Lett.}~{74B}(1978), 64.
\bibitem{helgas} S.~Helgason:
                 Academic Press, New York 1978.
\bibitem{BJO} E.~Bergshoeff, B.~Janssen, T.~Ort\'{\i}n:
              {\it Class.~Quantum Grav.}~{\bf 13}(1996), 321.
\bibitem{buscher} T.~Buscher:
                  {\it Phys.~Lett.}~{\bf 159B}(1985), 127;
                  {\bf 194B}(1987), 59;
                  {\bf 201B}(1988), 466.
\bibitem{kall2} R.~Kallosh, T.~Ort\'{\i}n:
                {\it Phys.~Rev.}~{\bf D48}(1993), 742.
\bibitem{bdw} B.S.~de Witt:
              {\it Phys.~Rep.}~{\bf 19C}(1975), 295.\\
              Ya.B.~Zel'dovich:
              {\it JETP Lett.}~{\bf 14}(1971), 180.\\
              A.A.~Starobinsky:
              {\it Sov.~Phys.~JETP} {\bf 37}(1973), 281.
\bibitem{kn:FHM} J.A.H.~Futterman, F.A.~Handler, R.A.~Matzner:
                 {\sl Scattering from Black Holes},
                 Cambridge University Press, 1988.
\bibitem{kn:BMPV} J.C. Breckenridge, R.C. Myers, A.W. Peet,
                  C. Vafa:
                  {\it Phys.~Lett.} {\bf B391}(1997), 93.
\end{thebibliography}
\end{document}